# Comprehensive Review on the Control of Heat Pumps for Energy Flexibility in Distribution Networks

Gustavo L. Aschidamini[1], Mina Pavlovic[1], Bradley A. Reinholz[2], Malcolm S. Metcalfe[2],
Taco Niet[1], and Mariana Resener[1]

[1]School of Sustainable Energy Engineering, Faculty of Applied Sciences, Simon Fraser University, Surrey, BC
V3T 0N1, Canada

[2]Generac Power Systems Inc., Waukesha, WI, USA

Email: gla113@sfu.ca

*Abstract*—Decarbonization plans promote the transition to heat pumps (HPs), creating new opportunities for their energy flexibility in demand response programs, solar photovoltaic integration and optimization of distribution networks. This paper reviews scheduling-based and real-time optimization methods for controlling HPs with a focus on energy flexibility in distribution networks. Scheduling-based methods fall into two categories: rule-based controllers (RBCs), which rely on predefined control rules without explicitly seeking optimal solutions, and optimization models, which are designed to determine the optimal scheduling of operations. Real-time optimization is achieved through model predictive control (MPC), which relies on a predictive model to optimize decisions over a time horizon, and reinforcement learning (RL), which takes a model-free approach by learning optimal strategies through direct interaction with the environment. The paper also examines studies on the impact of HPs on distribution networks, particularly those leveraging energy flexibility strategies. Key takeaways suggest the need to validate control strategies for extreme cold-weather regions that require backup heaters, as well as develop approaches designed for demand charge schemes that integrate HPs with other controllable loads. From a grid impact assessment perspective, studies have focused primarily on RBCs for providing energy flexibility through HP operation, without addressing more advanced methods such as real-time optimization using MPC or RL-based algorithms. Incorporating these advanced control strategies could help identify key limitations, including the impact of varying user participation levels and the cost-benefit trade-offs associated with their implementation.

## I. Introduction

Decarbonization plans include improving the energy efficiency of homes and buildings, promoting the transition to electric heat pumps (HPs) and creating new opportunities for their energy flexibility. Policies supporting this transition include rebates and incentives aimed at reducing reliance on natural gas [1] [2]. HPs, which replace natural gas-based heating systems, are used for space heating (SH), space cooling (SC), and domestic hot water (DHW) heating. Their widespread adoption highlights the importance of advanced control strategies that extend beyond maintaining thermal comfort to optimizing energy flexibility and interaction with the distribution grid.

The electrification of heating and cooling with HPs presents significant challenges to existing power distribution systems, which are not yet equipped to handle their widespread adoption. Despite the clear energy efficiency advantages of HPs, which are 2 to 4 times more efficient than electric-resistance heating systems [3, 4], their large-scale deployment can lead to increased network congestion and voltage issues [5, 6]. Therefore, to mitigate these challenges and enable a smoother integration of HPs into power distribution systems, possible solutions include network expansion and modernization [7], voltage regulation [8], and demand-side management (DSM) [8–10].

Given the significant impact of HPs on distribution networks, DSM strategies offer a cost-effective approach to mitigating voltage problems and network congestion in distribution networks with high HP adoption, reducing the reliance on expensive infrastructure upgrades. [7]. Traditional solutions involve upgrading distribution transformers to those with a higher power rating, replacing existing distribution cables with higher-capacity ones, and increasing feeder voltage, but these require significant investments and widespread implementation across multiple distribution feeders. Looking at passive distribution networks, voltage regulation is commonly performed with capacitor banks, voltage regulators, and on-load tap changers [11, 12]. However, solely relying on these legacy devices may not be sufficient to support the existing distribution network during the shift towards electrification [13]. Due to their electro-mechanical nature, these devices lack the speed required for managing the uncertainties in power demand within active distribution networks [11]. Instead, grid-support solutions incorporate customer equipment for reactive power compensation using solar photovoltaic (PV) and battery inverters [14], as well as DSM through distributed energy resources (DERs), such as scheduling electric-vehicle (EV) charging to shift demand to off-peak periods [15, 16]. By leveraging DSM approaches, distribution networks can better accommodate HP adoption while mitigating infrastructure reinforcements.

Control strategies for HPs are required for optimizing energy use through energy flexibility approaches, including price-based, incentive-based, and market-based mechanisms [17, 18]. These mechanisms can be categorized into direct control, where the aggregator or distribution system operator (DSO) has control access to heating systems, and indirect control, where adjustments occur without external intervention, such as through demand response (DR) programs [18, 19]. A congestion management mechanism with direct control is discussed in [19], with the possibility of curtailing residential loads to avoid blackouts. Additionally, price-based mechanisms, such as time-of-use (TOU) tariffs, differentiate between peak and off-peak periods, enabling customers to lower electricity expenses while potentially reducing peak demand [9]. For instance, in [10], HPs are locally controlled within a TOU tariff program, shifting heating-related power demand away from high-cost periods while minimizing impacts on thermal comfort. Therefore, implementing effective HP control strategies is essential for managing energy use and enabling the adoption of DSM strategies by customers.

Control strategies for HP systems vary in complexity, from simple rule-based controllers (RBCs) to advanced algorithms like reinforcement learning (RL). RBCs are simple strategies that use predefined rules to control the indoor temperature or hot water tank temperature [20]. These do not aim for an optimal solution but instead operate deterministically and with low computational complexity, lacking the flexibility to handle uncertainties. In contrast, mathematical optimization models, such as mixed-integer nonlinear programming (MINLP) [21] and mixed-integer linear programming (MILP) [22], seek optimal solutions, but involve higher computational complexity. Model predictive control (MPC) is used for real-time optimization, addressing forecast uncertainties and generally outperforming RBC [23]. RL's free approach allows it to operate without needing a predefined system model, making it ideal for real-time optimization. Consequently, RL algorithms have been increasingly used in heating systems for DR programs, owing to their low computational time requirements for real-time operations [24]. These approaches demonstrate the evolving complexity of HP control strategies, underscoring the need for a comprehensive review of their role in DR and distribution network management. Therefore, this review explores mitigation strategies for enhancing energy flexibility and investigates the impact of HP adoption on distribution networks, emphasizing the importance of understanding these effects for effective network management.

Although many reviews focus on specific areas within DSM involving DERs, few address the operation and control of HPs. For instance, [18] reviews coordination strategies for DERs connected at the edge of distribution networks, exploring the control access of units, objectives, and coordination mechanisms. A review on RL for DR was presented in [25], showing applications to control EVs, batteries, PV systems, and heating, ventilation, and air conditioning (HVAC). Additionally, [26] provided a review on DR and load scheduling strategies including those incorporating advanced information and communication technologies.

Some reviews specifically focus on DR programs and their mechanisms for network management. For example, [27] reviews pricing-based DR programs, in which the consumption behavior of customers is affected by varying electricity prices, with the objective of reducing electricity expenses. Additionally, the reviews in [28] and [19] explore characteristics of strategies for congestion management in distribution networks.

A few review papers concentrate on the control of HPs. In [29], HP control in distribution networks is examined and classified based on its applications to technical grid aspects, renewable energy integration, and electricity price considerations. Model-based analyses for integrating residential HPs and thermal energy storage (TES) systems with the aim of facilitating renewable energy integration were reviewed in [30]. A review of rule-based control strategies and MPC strategies for HPs providing flexibility to the grid is provided in [31]. The research in [32] investigates the use of TES systems for providing demand flexibility. Metaheuristic methods for controlling HPs and TES units were reviewed in [33], with a focus on grid-edge strategies. Despite these advancements, there remains a need for a comprehensive review that focuses specifically on the control of HPs and their role in managing distribution network congestion.

This paper addresses these gaps by examining the control strategies for HPs and their contribution to energy flexibility within distribution networks, through the following objectives:

- Examine the main control strategies involving HPs, focusing on their architecture and how they contribute to energy flexibility;
- Analyze the main methods used to control HPs within DR programs for both direct and indirect control approaches, examining energy flexibility, devices involved, and objectives;
- Highlight relevant limitations of control strategies for HPs from the existing literature, leading to recommendations for future research;
- Define limitations of grid impact assessment studies of HPs on distribution networks.

*A. Contributions*

In response to the challenges of HP integration and distribution network congestion outlined earlier, we offer a handy resource that provides a concise summary of the most relevant publications on models used for HP control. The main contributions of this paper are as follows:

- A background on the architecture and main methods in the context of parameters for HP control;
- Building on this background, the second contribution reviews rule-based, optimization, and RL-based methods for controlling HPs and auxiliary heating systems;
- An analysis of how these systems are coordinated with other devices, such as solar PV panels and battery energy storage systems (BESS);
- A review of grid impact assessment studies of HPs on distribution networks;
- Recommendations for future research in control strategies of HPs and auxiliary heating systems and grid impact assessment studies.

## B. Paper Organization

This paper has been organized into the following sections: Section II presents the methodology of this literature review. Section III provides a background on the operation of HPs for providing energy flexibility and analyses methods for the operation of HPs. Section IV reviews RBCs; optimization models, MPC methods; and RL-based methods. Section V presents a review of grid impact assessment studies of HPs on distribution networks. Finally, we provide conclusions and recommendations for future work in Section VI.

## II. METHODOLOGY OF THIS LITERATURE REVIEW

This section outlines the methodology for finding, analyzing, and synthesizing the relevant literature.

### A. Literature Search Process

The objective of this paper is to explore HP control schemes and their impact on distribution networks. Scopus [34] database was selected due to its comprehensive coverage of peer-reviewed literature in the fields of energy management and control strategies. A search in the Scopus database using the keywords "heat" AND "pumps" AND "energy" AND "management" resulted in 2,197 papers published between 2014 and 2024. Additionally, a search with the keywords "heat" AND "pumps" AND "demand" AND "response" resulted in 677 papers. Fig. 1 illustrates the number of papers published per year from these two searches. An upward trend in the number of papers published by year suggests an increasing research focus on HPs and their role in DSM. The combined results from these searches served as the database for this literature review.

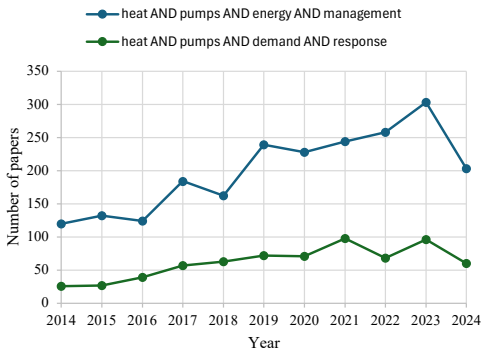

Fig. 1: Papers published by year for a search in the Scopus database.

### B. Selection Criteria

Following the initial search, we selected 70 papers using a binary include/exclude decision based on their focus on HP control strategies for energy management, as well as their integration with other technologies such as BESS and solar PV systems. Note that papers that include the control of district heating systems, but not HPs, were not included in this literature review. For a review of DSM with district heating, refer to [35].

### C. Methodological Analysis

To address the research gap in HP control strategies, we followed a structured methodological analysis. For each selected paper, we identified (i) the types of devices controlled by or included in the simulations and (ii) the methodological approach used, whether it followed a rule-based approach, an optimization model (such as MINLP and MILP), an MPC, or an RL-based method.

### D. Comparison and Evaluation

*1) Identification of Control Strategies:* Building on the methodological approach proposed in each paper, we identified the characteristics of the heating systems associated with each method. This includes (i) the type of HP based on the heat source, such as air-source HPs (ASHPs) or ground-source HPs (GSHPs); (ii) the applications of HPs, including DHW heating, SH, SC, or a combination of these; and (iii) the backup heating system, if considered, such as an electric resistance heater (ERH) or a gas boiler (GB).

*2) Assessment of Methodological Approaches:* We analyzed the management approaches, such as local energy management system (EMS) control or centralized control of multiple HPs, and the energy flexibility focus, considering whether it follows electricity costs, integrates PV generation, or focuses on network goals. The objectives of each method were identified, such as reducing electricity expenses, maximizing PV self-consumption, and mitigating network congestion and overvoltage.

*3) Comparison and Highlighting of Limitations:* These patterns provided an initial comparison between the methods. It is followed by highlighting limitations and comparing strategies found in the existing literature.

### E. Grid Impact Studies Search Process

The grid impact studies on HPs in distribution networks were selected through a different search process.

*1) Literature Search Process:* We searched journal articles and conference papers that examine the impact of HPs on distribution networks. A search in the Scopus [34] database using the keywords "heat" AND "pumps" AND ("distribution" OR "low-voltage") AND ("grids" OR "networks") identified 1,111 papers published between 2014 and 2024.

*2) Selection Criteria:* Following this initial search, we selected 20 papers based on an inclusion/exclusion criterion that focused on whether the study incorporated HPs in a grid impact assessment. Therefore, only papers specifically examining HPs were included, while those focusing solely on electric water heaters (EWH) or district heating systems were excluded.

*3) Comparison and Evaluation:* For comparison and evaluation of the selected papers, we examined the devices used, including the type of HP and the integration of other systems such as solar PV, EVs, or other installations. We also assessed the use of grid violation metrics, such as demand calculations, voltage unbalance factors, and overloading conditions. Lastly, where applicable, we evaluated the proposed solutions for mitigating grid violations through the energy flexibility of HP power demand.

## III. TECHNICAL BACKGROUND ON HEAT PUMP OPERATION FOR ENERGY FLEXIBILITY

In this section, we present the theoretical background on the operation of HPs for providing energy flexibility, including the architecture of their control, strategies, and control methods.

### A. Architecture for Heat Pump Control

The control architecture of flexible loads, including HPs, is predominantly designed around DR strategies. These strategies aim to modify electricity consumption patterns of demand-side resources, either through voluntary or mandatory participation contracts [36]. DR programs are commonly classified into price-based and incentive-based mechanisms, where incentive-based strategies are further sub-categorized into financial compensations or market-based approaches.

Understanding these mechanisms is critical because they directly impact how HP electricity consumption or demand is billed. Among DR strategies, price-based DR utilizes variable electricity prices to motivate end-users to adjust their energy consumption patterns. Fig. 2 presents four price-based DR mechanisms. TOU tariffs, as illustrated in Fig. 2a, commonly introduce higher electric tariffs during peak demand periods, incentivizing users to shift their energy consumption to off-peak times. These tariffs are predictable, offering protection from unexpected electricity price changes. In contrast, real-time pricing (RTP), as shown in Fig. 2b, allows tariffs to fluctuate based on market conditions, resulting in more volatile prices and requiring smart metering to track consumption. Together, these mechanisms encourage customers to align their energy usage with network conditions.

Beyond TOU and RTP, other price-based DR mechanisms, such as critical peak pricing (CPP) and demand charges, further incentivize shifts in energy consumption to reduce peak loads. CPP, as shown in Fig. 2c, raises electricity prices during periods of grid congestion or technical issues, prompting users to lower demand during critical events. Another mechanism, demand charges, as illustrated in Fig. 2d, is based on peak energy usage rather than total energy consumption and is typically applied to commercial and industrial customers to manage their higher peak loads. Collectively, CPP and demand charges introduce more variability in electricity billing compared to TOU and RTP, which could significantly impact overall electricity costs if the heating systems operate during high-cost periods.

DR strategies based on signals, such as price-based and market-based approaches, are examples of indirect control methods that provide flexibility to end-users without direct remote dispatch [18, 29]. In price-based approaches, end-users manage their own heating systems settings, while market-based approaches aim to balance power supply and demand through a structure involving systems operators and market participants [37]. These include demand-bidding, capacity, and ancillary services markets, which offer incentive-based DR options through payments to participants [38]. While these programs allow customers to participate in DR, they still rely on indirect control, providing flexibility without direct remote dispatch from a DSO or aggregator. Indirect control strategies depend on voluntary participation, meaning actual changes in consumption are not guaranteed. Furthermore, price-based strategies may trigger rebound effects, where energy use that was reduced or shifted during a DR event increases afterward.

Direct load control (DLC) and curtailable load programs are alternative DR strategies that offer immediate and guaranteed control over energy demand during network congestion events. DLC provides an alternative approach, where end-users receive financial compensation for granting control over their equipment, like heating systems, to a DSO or aggregator [19]. Similarly, in curtailable load programs, the DSO or aggregator gains access to specific customer loads that can be turned off as needed, offering more immediate and guaranteed control over energy demand during network congestion events.

### B. Control Strategies for Energy Flexibility of Heat Pumps

Control strategies for energy flexibility through HP operation can be implemented for both HP operating for SH and DHW heating functions.

*1) Energy Management System:* To coordinate devices within a building, an EMS plays a crucial role, particularly when integrating technologies like HPs for SH and DHW heating applications. Fig. 3 illustrates a typical ASHP system, which is often integrated into EMS for these applications. ASHPs operate by transferring heat from the outside air through the processes of evaporation, compression, condensation, and expansion. The ASHP can function for SH or DHW heating applications. For DHW heating, tap water is heated in the hot water tank. In extreme cold conditions, when the ASHP's efficiency drops, the backup heating system is activated to ensure thermal comfort by maintaining indoor ambient temperature and hot water supply. As for SH, Fig. 3 shows two configurations: the air-to-air, which uses an inside coil to distribute air throughout the building, and air-to-water, which transfers heat to a set of radiators or in-floor heating systems, with the option of using a buffer tank. Additionally, the heating system includes backup ERH for both SH and DHW heating. The primary function of the HP power demand controller is to maintain thermal comfort while maximizing energy efficiency [39]. The EMS is responsible for controlling and optimizing the operation of these heating systems, particularly in coordination with DR programs.

Alternative HP configurations, such as GSHPs and dual-source HPs (DSHPs), offer distinct advantages in maintaining efficient heating performance, which can be leveraged by the EMS to ensure thermal comfort while providing energy flexibility. GSHPs generally have more a consistent coefficient of performance (COP) over the year because they extract heat from the ground, which maintains a relatively constant temperature [40]. Additionally, DSHPs combine two heat sources, such as water from a thermal energy storage system and air [41].

*2) Space Heating:* One approach for providing energy flexibility through the SH functionality involves preheating the indoor ambient temperature, enabling the shifting of power demand to off-peak periods [42]. This approach can slightly increase the indoor ambient temperature above the target level without compromising thermal comfort. The thermal inertia of buildings is leveraged as a form of thermal storage, while improved insulation reduces thermal losses. For instance, [43] demonstrated that preheating the indoor temperature 1 to 3 hours before the expected peak demand periods can contribute to reducing the peak demand. In air-to-water HPs, SH buffer tanks are sized to reduce the "turn on and off" short cycles of HP with the aim of extending the compressor's lifespan. These can be combined with a solar thermal collector (STC) for heating the water in the buffer tank. In this heating system configuration, preheating is in fact heating the water in the buffer tank at off-peak demand times.

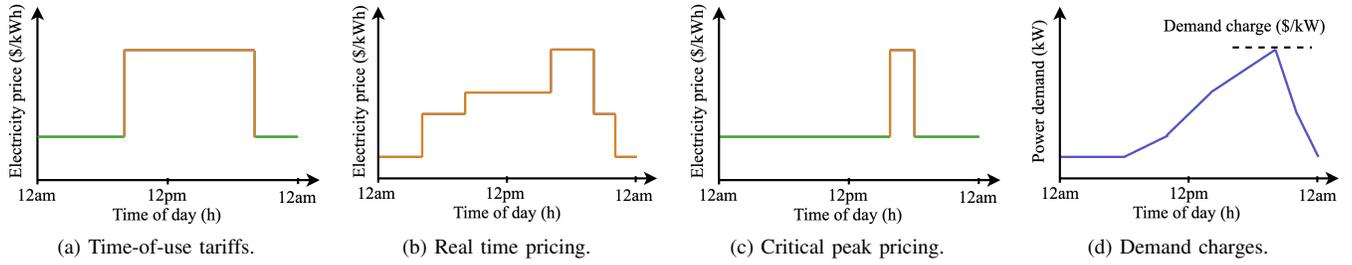

(a) Time-of-use tariffs.    (b) Real time pricing.    (c) Critical peak pricing.    (d) Demand charges.

Fig. 2: Price-based DR mechanisms.

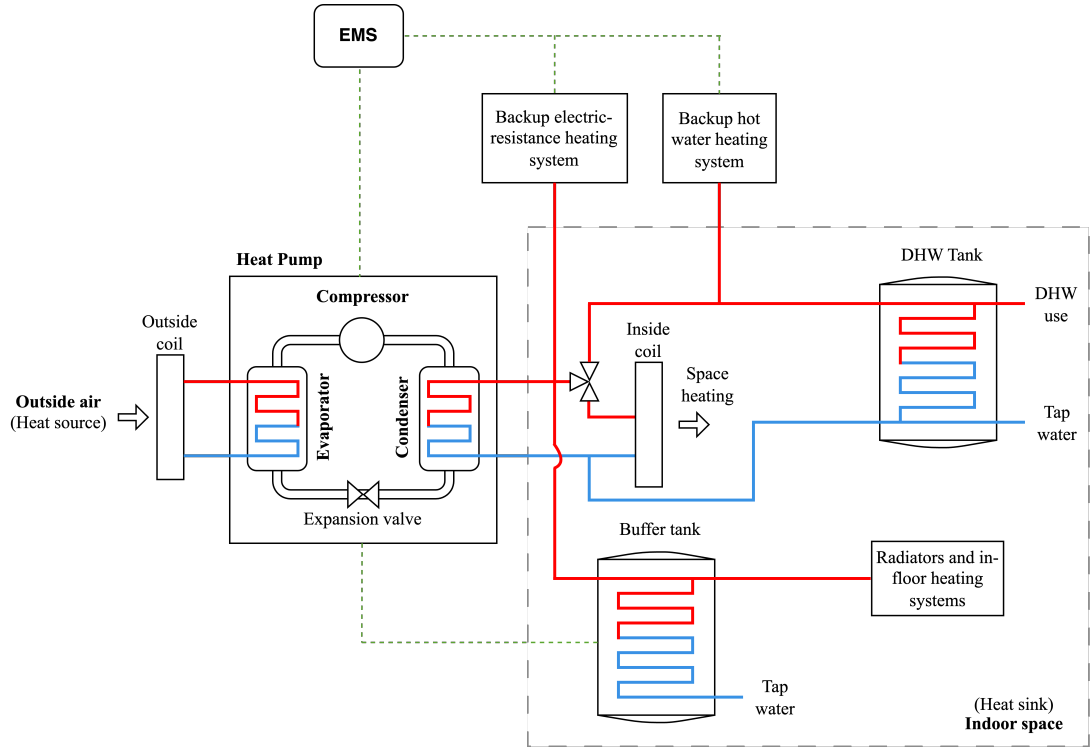

Fig. 3: Diagram of ASHP for SH and DHW heating use with a backup heating system.

In addition to preheating, another strategy involves reducing the indoor ambient temperature during peak demand times by adjusting the target temperature. This indoor temperature setback can further decrease heating power demand, but it may affect thermal comfort, as the settings might not align with user preferences for indoor ambient temperature.

While the previously discussed strategies can be imple- mented in indirect load control, another alternative is to curtail the operation of heating systems, which has been studied in [28]. In this case, however, the impact on power demand and indoor ambient temperature should be investigated as it may impact thermal comfort, especially in extreme cold weather scenarios.

*3) Domestic Hot Water Heating:* Building on the discussion of SH control for energy flexibility, preheating the hot water tank serves as an additional form of TES [44]. In this approach, the water temperature in the tank is increased during off-peak demand periods, such as those associated with TOU tariff programs. This allows for a shift in power demand without significantly affecting thermal comfort [45]. However, thermal losses must be taken into account to ensure that thermal comfort is maintained during peak demand periods. When the HP has functions for both DHW heating and SH, a control strategy must balance the priority between heating the water tank and the indoor ambient temperature [46], therefore taking into account the thermal comfort levels for both hot water and indoor ambient temperature.

## C. Methods for Energy Flexibility Through Heat Pump Operation

We categorized the methods utilized by the selected papers for controlling HPs and other devices into (i) RBCs; (ii) optimization models based on scheduling (e.g. MINLP, MILP); (iii) MPC; and (iv) RL-based methods. The following sections discuss these methods in the context of HP operation for energy flexibility.

Table I summarizes the main control methods for HP operation, categorizing them by complexity, computational requirements, and their suitability for different energy flexibility objectives. This overview highlights the strengths, limitations, and typical applications of each approach.

As shown in Table I, RBCs offer simplicity and low computational demand but lack adaptability to dynamic conditions. Optimization-based methods provide precise scheduling but require accurate forecasts and are less flexible in real-time. In contrast, MPC balances flexibility with computational effort by continuously updating schedules based on forecasts, while RL methods stand out for their adaptability but demand significant computational resources during training.

The following subsections examine each control method in greater detail, assessing their specific applications, computational demands, and contributions to energy flexibility.

*1) Rule-based Operation:* RBCs represent a simpler category of control methods due to their low computational effort when implemented solely based on measurements, without the need for power demand forecasting [47]. These controllers operate according to a predefined set of rules that determine the "On" or "Off" state of HP for SH and DHW heating, as well as managing variable-speed compressors and their output power. Hysteresis is commonly employed in these systems for maintaining temperature within user-defined preferences while minimizing frequent cycling [50]. Control decisions are informed by temperature sensors placed within water tanks and monitoring indoor ambient temperature, and objectives such as electricity prices are integrated into the rule set.

While rule-based methods do not demand significant computational resources, they also do not aim for an optimal solution [51]. In contrast, HP control strategies that utilize optimization models and RL-based methods typically demonstrate superior performance compared to RBCs [52]. Optimization models can be categorized into optimal scheduling and MPC, which are discussed in the following sections.

*2) Optimization Models Based on Scheduling:* Optimization models based on scheduling aim to define optimal heating and cooling control actions, typically using mathematical programming techniques [53]. These models are built around an objective function, which commonly minimizes operational costs [21]. The objective function often consists of two terms: one representing the main goal (e.g., electricity costs), and another penalty term that introduces a cost for constraint violations [46]. This is known as a soft constraint formulation, where the model does not strictly enforce constraints. Additionally, the model incorporates parameters such as thermal comfort limits, and the availability of energy and heating resources.

The optimal minimization of electricity costs through the control of HPs and other devices is often formulated as a MILP model [22]. However, if the operation of power distribution networks is also considered, a MINLP formulation is normally used due to the inclusion of power flow equations [54].

*3) Model Predictive Control:* Moving beyond these optimal scheduling approaches, MPC is commonly used to represent dynamic systems and focuses on real-time optimization [55]. Instead of setting predefined schedules like mathematical optimization models, MPC continuously adjusts control actions by solving an optimization model for a future time horizon, allowing it to act to real-time changes in the system.

MPC methods vary in both complexity and their representation of uncertainties, ranging from classic MPC to adaptive MPC (AMPC). Classic MPC is suitable when it is known that the model is accurate, like in deterministic problems. Remedying this, robust MPC (RMPC) extends the features of classic MPC by dealing with uncertainties for the worst-case scenario [56]. Stochastic MPC (SMPC) also deals with uncertainties, however, through probabilistic distributions [57]. In this approach, the controller optimizes the cost function while ensuring that the constraints are met with a specified probability. AMPC adapts to system dynamics in real-time operation over the specified run time [58]. MPC formulations that incorporate uncertainties and external disturbances tend to be more robust than classic MPC, underscoring the complexity of managing uncertainty in real-time HP systems.

*4) Reinforcement Learning:* While MPC requires building a model of the system's dynamics to predict future states and optimize control actions, most RL algorithms follow a model-free approach. In this approach, the agent learns through trial and error by interacting with the environment, allowing it to make sequential decisions without requiring an explicit model of the system [59]. However, some RL algorithms adopt a model-based approach, where the agent learns a model of the environment's dynamics to guide the decision-making process [60]. Regardless of whether the approach is model-free or model-based, RL typically frames these sequential decision-making problems within the Markov decision process (MDP) framework, which serves as the foundation for how RL agents learn optimal control policies.

The MDP is commonly used to model these sequential decision-making problems [59]. Fig. 4 presents a diagram of MDP within the context of energy management with thermostatically controlled loads (TCL) [59]. The MDP consists of an agent ($A_t$), an environment, a reward ($R_t$), a set of actions ($A_t$), and a set of states ($S_t$). In this context, the agent is the EMS, which acts as a local or centralized controller that interacts with the environment, consisting of components such as HPs, DHW tanks, buffer tanks, and backup heating systems. The environment is often simulated using a thermal dynamics model when direct interaction with actual equipment is not feasible. The agent's actions typically involve setting the target hot water temperature, indoor ambient air temperature, or the state of HP and backup heating systems. The states, retrieved from the environment, are related to variables such as outdoor air temperature, COP of the HP, and the state of charge in the hot water tanks. A reward function, typically composed of objectives like minimizing electricity costs and penalizing thermal discomfort, serves as feedback for the agent. The MDP problem is then solved by RL algorithms, allowing the agent to learn optimal control strategies.

TABLE I: Summary of control methods for HP operation.

| Control method | Computational complexity | Real-time adaptability | Performance (flexibility and optimization) |
| --- | --- | --- | --- |
| Rule-based control | Low [43] (higher for methods that do require heating load forecast [47]) | High (reacts instantly but lacks optimization) | Limited (predefined logic, not optimal) [48] |
| Optimization based on scheduling (MILP, MINLP, etc.) | High (especially for MINLP [21]) | Low (offline scheduling, not adaptable in real-time) | High for predefined scenarios but inflexible to real-time changes [46] |
| Model predictive control | Medium to high (MINLP-based usually require higher computational time compared to MILP-based MPC) [44] | Medium to high (adaptability depends on update interval and model accuracy) | High (adaptive but requires accurate models) [44] |
| Reinforcement learning | Very high (training phase [49]) | High (adaptive, but continuous learning depends on the method) | Very high (adaptive, learns optimal policies) [44] |

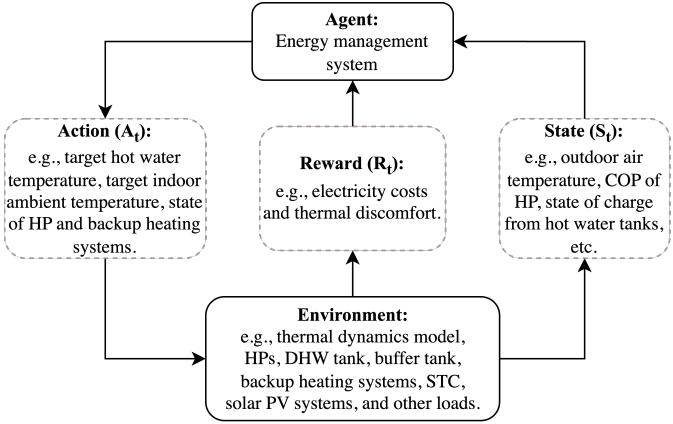

Fig. 4: Example of MDP in the context of energy management with TCL.

In terms of action representation, the agent's action space can be either discrete or continuous. Continuous states enable the modeling of variable-speed or modulating compressors, which have continuous output power, whereas discrete states are appropriate for binary control scenarios. States can also incorporate multiple discrete levels for implementing the actions of variable-speed compressors.

The architecture of RL algorithms can be categorized into three main types: value-based, policy-based, and actor-critic methods [59]. Value-based methods focus on learning a value-function for state-action pairs or states, which guides the decision towards actions that maximize rewards. In contrast, policy-based methods explicitly optimize the policy that the agent uses to choose actions. Actor-critic methods combine both approaches; the actor learns to select actions that maximize the reward function, while the critic assesses these actions by calculating the value of each state-action pair.

Building on this architectural foundation, model-free value-based methods aim to learn optimal action selection through the value function. Classic tabular Q-learning utilizes Q-tables to store values of state-action combinations, which is only applicable for discrete state and action spaces [59]. In contrast, deep RL allows continuous state spaces through advanced function approximations [61]. Deep Q-networks (DQN) [61] utilize neural networks to approximate Q-values in high-dimensional and continuous state spaces. Double Q-learning (DQL) [62] utilizes two separate Q-tables or neural networks to mitigate the overestimation of action values present in DQN [63]. Collectively, these methods represent the core approaches in model-free value-based RL.

Policy-based methods in RL aim to improve training stability and performance. Trust region policy optimization (TRPO) is designed to ensure that policy updates remain within a specified trust region to maintain stability during training [64]. Proximal policy optimization (PPO) improves upon this by utilizing neural networks to simultaneously approximate both the policy and value functions, enhancing stability and achieving better performance [65].

Actor-critic methods combine the benefits of both value-based and policy-based approaches. The deep deterministic policy gradient (DDPG) [66] is particularly suited for continuous action space, while the twin delayed deep deterministic policy gradient (TD3) enhances DDPG by utilizing two Q-networks to mitigate overestimation issues. However, TD3 introduces a delay for policy updates and adds noise to actions to encourage exploration [67]. The soft actor-critic (SAC) incorporates an entropy term in the reward function, promoting exploration and allowing the agent to try a more diverse set of actions, but this can occasionally lead to unsafe actions in real-world applications [68].

Unlike online algorithms, offline or batch RL trains on a predetermined historical dataset rather than relying on real-time interaction with the environment. For example, fitted Q-iteration (FQI) [69] [70] learns from historical data without requiring additional integration with the environment during training, making the training process safer.

While the aforementioned RL methods are model-free, model-based approaches, such as Monte Carlo tree search (MCTS) [60], provide an alternative structure for control. MCTS allows the agent to learn a model of the environment's dynamics, guiding decision-making through an online iterative search process that explores actions and states. Together, these RL strategies offer a balance between exploration and safety in managing HP operations, both online and offline.

This classification not only allows us to understand how each control method operates independently but also reveals the potential for hybrid approaches that combine the strengths of different methods.

## IV. REVIEW OF HP CONTROL METHODS FOR ENERGY MANAGEMENT

To organize the methods utilized by the selected papers for controlling HPs, we categorized them into RBCs, optimization models based on scheduling; MPC, and RL-based methods. The following sections discuss these methods.

### A. Rule-based Methods

The discussion of the selected papers focusing on rule-based methods is supported by Table II, which provides a summary of these methods in HP operation, categorized by DR schemes, renewable energy integration, and network objectives. Although energy flexibility was sorted into these three categories, some papers fall into more than one category.

One approach to providing energy flexibility through HPs and other devices includes leveraging DR programs with local device control. For instance, [20] proposed a controller for a GSHP, while locally managing a solar PV system and BESS in a dynamic pricing market. This demonstrated that coordinating residential devices reduces electricity costs and increases self-consumption. A study in [50] developed a predictive RBC for an ASHP with SH and DHW functions with an STC that increases temperature, reducing heating electricity costs, peak-period power demand, and carbon emissions. This control method was also applied to auxiliary electric heaters for DHW and SH, whose impacts are particularly relevant in cold-weather regions. While predictive RBCs rely on forecasts of future conditions, traditional RBCs operate only on predefined rules. In [74], a solar PV system and ASHP are coordinated through a DR program, optimizing local grid electricity prices and solar feed-in tariffs while effectively lowering annual power demand and increasing self-consumption. Additionally, [75] introduced a rule-based method using regression models to control an HP for DHW heating alongside a solar PV system and BESS, considering TOU tariffs to reduce electricity expenses. By predicting electricity consumption and solar power generation, the controller can take proactive actions. While these methods control flexible loads within DR programs, further expansion in this field focuses on coordinating the energy management of distributed generation (DG).

Other approaches focus on increasing energy flexibility without explicitly relying on DR programs. Instead, they aim to enhance self-consumption through the coordination of HPs and solar PV systems. For example, [73] proposed a coordination method with ASHPs, resulting in an additional peak-shaving function, thereby benefiting the distribution network. In [8], a centralized control method for multiple devices is proposed, including HPs, EVs with vehicle-to-grid (V2G) capabilities, solar PV systems, and BESS, which integrated network objectives by aiming to increase grid hosting capacity. Increased self-consumption offers clear advantages for the customer, and coordinating multiple devices can further enhance energy management, leading to explicit strategies that align with distribution network objectives.

Looking into local control methods, they focus on directly integrating network objectives into device control using local, centralized, or decentralized approaches, without the need to rely on DR programs or coordination with residential DG. For example, [71] introduces a droop control strategy for locally managing an ASHP for DHW and SH, with the aim of mitigating overvoltage in low-voltage (LV) distribution networks. Controlling the ASHP in this manner helped mitigate the impact of solar PV generation on the network. The work in [72] locally manages an HP for DHW and an EV charger to alleviate grid congestion and reduce voltage deviations at the distribution network's point of connection. Building upon this, [45] locally controls an ASHP for DHW to perform peak shaving, while simultaneously managing a backup ERH system. The method curtailed power demand for two hours of DHW heating, which provided energy flexibility without affecting thermal comfort. A DLC strategy [47] utilizes an ASHP for SH as another peak-shaving solution. The effectiveness of this strategy was enhanced through the addition of higher accuracy power load predictions within the predictive RBC. While these methods focus on local energy management for individual end-users, the coordination of devices across multiple buildings can leverage network objectives on a broader range of distribution network nodes.

Centralized and decentralized approaches to controlling HPs for demand reduction are explored, along with coordination strategies to enhance their effectiveness. Analyzing the use of multi-building device coordination, [40] implements centralized control of multiple ASHPs and GSHPs for peak shaving, showing that smart grid integration of HPs can offer greater flexibility than simple curtailment approaches. Additionally, [43] applies a decentralized approach to control ASHPs for DHW and SH, which showed that using multiple temperature setbacks was more effective in reducing heating peak demand, compared to preheating. However, while the method assessed the benefits of decentralized control of HPs, their coordination was not established. A more flexible approach to mitigating technical challenges on a distribution network level lies in rule-based methods that align customer device coordination with network objectives.

While RBCs can coordinate the operation of HPs with other devices within DR programs, they may not provide the optimal scheduling for heating systems that hope to achieve network objectives or certain levels of solar PV integration. This limitation highlights the need for mathematical optimization methods, which play a crucial role in defining these optimal control signals. Such methods include models for HP scheduling, as well as MPC approaches that optimize in real-time. Through the use of these optimization techniques, EMS can improve the coordination of local or multiple devices, addressing broader distribution network objectives while benefiting end-users.

### B. Optimization Models Based on Scheduling

This section explores non-linear and linear optimization models for HP scheduling control. To facilitate a comprehensive discussion, Table III categorizes these approaches, focusing on DR schemes, solar PV integration, and network objectives. The table also highlights the key differences and similarities among these models.

The selected schedule-based optimization models primarily focus on coordinating the operation of HPs and other devices for customers enrolled in DR programs. Some of these models concentrate on customer loads without explicitly incorporating DG into their formulation. For example, an optimization model proposed in [82] combines an artificial neural network (ANN) for locally controlling a GSHP used for SH and SC under TOU tariffs, aiming to minimize electricity costs by predicting heating demand. In [83], the preheating coordination of HPs for SH using a local and central controller based on an optimization model is proposed, reducing both electricity and generation costs by accounting for marginal electricity prices. Additionally, [85] introduced a linear programming (LP) model for local control of an ASHP for SH and SC, with the goal of minimizing electricity costs under RTP. While these models effectively manage loads under DR programs, they do not explicitly integrate solar PV generation into the coordination process.

TABLE II: Summary of rule-based control methods for HP operation.

| Ref. | Method | Devices | HP use | HP compressor speed | | Backup system | Management approach | Energy flexibility focus | | | Objective |
|---|---|---|---|---|---|---|---|---|---|---|---|
| | | | | Fixed | Variable | | | DR | PV | Network | |
| [71] | Droop control | ASHP and TES | DHW and SH | - | ✓ | - | Local control | - | - | ✓ | Mitigate overvoltage |
| [40] | Rule-based | ASHPs and GSHPs | DHW and SH | ✓ | - | ERH | Centralized control of multiple HPs | - | - | ✓ | Peak shaving |
| [20] | Rule-based | GSHP, solar PV system, and BESS | DHW and SH | ✓ | - | - | Local control | Dynamic pricing market | ✓ | - | Reduce electricity expenses and increase self-consumption |
| [50] | Predictive rule-based | ASHP and STC | DHW and SH | ✓ | ✓ | ERH | Local control | Day-ahead spot pricing | - | - | Reduce energy costs for heating, reduce annual $CO_2$ emissions and reduce energy use for heating during peak hours |
| [72] | Rule-based | HP and EV charger | DHW | ✓ | - | - | Local control | - | - | ✓ | Mitigate grid congestion and voltage deviation |
| [73] | Rule-based | ASHP, TES, and solar PV system | DHW, SH, and SC | ✓ | - | - | Local control | - | ✓ | - | Peak shaving and increase energy self-consumption |
| [43] | Rule-based | ASHPs | DHW and SH | ✓ | - | - | Decentralized control | - | - | ✓ | Peak shaving |
| [74] | Rule-based | ASHP, TES, and solar PV system | DHW, SH, and SC | ✓ | | - | Local control | Local grid electricity prices and solar feed-in tariffs | ✓ | - | Reduce electricity demand and increase self-consumption |
| [8] | Rule-based | HPs, EVs with V2G capabilities, solar PV systems, and BESS | - | ✓ | - | - | Centralized control of multiple devices | - | ✓ | ✓ | Increase grid hosting capacity |
| [75] | Rule-based and regression models | HP, solar PV system, TES, and BESS | DHW | - | ✓ | - | Local control | TOU tariffs | ✓ | - | Reduce electricity expenses |
| [45] | Rule-based | ASHP | DHW | - | ✓ | ERH | Local control | - | - | ✓ | Peak shaving |
| [47] | Rule-based; load forecasting | GSHP | SH | - | - | - | DLC | - | - | ✓ | Peak shaving |

Some methods also enable the control of self-consumption by locally controlling an HP in combination with a solar PV system. For instance, [76] proposed a metaheuristic optimization model for controlling a WSHP, BESS, and solar PV system, with the aim of minimizing the cost of electricity and battery operation and maintenance. Similarly, [81] introduced a metaheuristic optimization model, solved using tabu search (TS), to control an HP for DHW heating, a solar PV system, and a BESS under TOU tariffs. Also aimed at minimizing electricity costs, this method allows for temporary disconnection from the grid, enabling local power supply for a smart house. Additionally, [48] developed an optimization model to manage a GSHP for SH with a solar PV system under TOU tariffs. The authors of [79] proposed a non-linear (NL), non-convex (NC) optimization model for controlling an ASHP for DHW heating and SH, a solar PV system, and a BESS under constant electricity prices, with the goal of maximizing the initial investment and minimizing operation and maintenance costs.

Further methods incorporate BESS and solar PV systems into local control strategies. The MINLP model proposed in [53] includes an STC for heating, aiming to provide TOU arbitrage and maximize self-consumption, while also responding to RTP. Another method in [21] formulated a MILP model that controls an HP and other flexible appliance loads under dynamic tariffs to minimize operation costs through a multi-objective approach, also incorporating a solar PV system. The operation of an HP, solar PV system, electric water boiler, and an EV charger is coordinated in [22] to integrate fixed, exchange price, and TOU tariffs for DR programs, demonstrating maximum savings on electricity costs.

While most papers include DR programs to harness the reduction of electricity expenses for customers, the use of the ancillary services market was also explored in the literature. The operation of an HP was coordinated with electric and biomass fire boilers in [38] using an optimization model, maximizing profit by participating in both the day-ahead electricity market and frequency regulation market.

TABLE III: Summary of optimization models for HP scheduling.

| Ref. | Model | Devices | HP use | HP compressor speed Fixed | HP compressor speed Variable | Backup system | Management approach | Energy flexibility focus DR | Energy flexibility focus PV | Energy flexibility focus Network | Objective |
|---|---|---|---|---|---|---|---|---|---|---|---|
| [76] | Heuristic | WSHP, TES, BESS, and solar PV system | DHW and SH | ✓ | - | - | Local control | TOU tariffs | ✓ | - | Minimize electricity costs and costs of battery operation and maintenance |
| [77] | Opt. | ASHPs, TES, EVs, home and community energy storage, flexible appliances, and solar PV system | DHW and SH | - | ✓ | ERH | Coordination of devices from a set of households, community energy system | TOU tariffs | ✓ | - | Minimize electricity costs and CO2 emissions |
| [78] | Heuristic | ASHPs, TES, and solar PV systems | DHW and SH | - | ✓ | - | Centralized and decentralized control systems for coordinating the operation of buildings | Day-ahead and intra-day market electricity prices | ✓ | - | Minimizing electricity costs and the surplus energy |
| [79] | NL, NC | ASHP, TES, BESS, and solar PV system | DHW and SH | - | ✓ | - | Local control | Constant electricity prices | ✓ | - | Minimize initial investment, cost of operation, maintenance costs and capital cost |
| [80] | MILP | ASHPs and TES | DHW and SH | - | ✓ | - | Coordination of a pool of HPs in a residential district | - | - | ✓ | Minimize power peak demand at the distribution network level |
| [81] | TS | HP, BESS, and solar PV system | DHW | ✓ | - | - | Local / DLC | TOU tariffs | ✓ | - | Minimize electricity costs |
| [82] | Opt.; ANN | GSHPs and TES | SH and SC | - | ✓ | - | Local control | TOU tariffs | - | - | Minimize electricity costs |
| [48] | Opt. | GSHP, TES, and solar PV system | SH | ✓ | - | - | Local control | TOU tariffs | ✓ | - | Minimize electricity costs |
| [83] | Opt. | HPs | SH | ✓ | - | - | Local and central controller for preheating coordination | Marginal electricity price | - | - | Minimize electricity costs and reduce generation cost |
| [53] | MILP | HP, TES, BESS, STC, and solar PV system | DHW and SH | - | ✓ | - | Local control | RTP and TOU tariffs | ✓ | - | Maximize self-consumption; and TOU arbitrage |
| [46] | LP | ASHPs | DHW and SH | ✓ | - | - | Individual peak shaving and central OPF control | - | - | ✓ | Minimize power losses |
| [38] | Opt. | HP, TES, electric boiler, and biomass fire boiler | SH | ✓ | - | ERH | Local control | Day-ahead electricity and frequency regulation markets | ✓ | - | Maximize profit |
| [84] | Convex | GSHPs and TES | SH and SC | ✓ | - | - | Local control and distributed coordination | - | - | ✓ | Minimize demand response deviation and increase COP of GSHPs |
| [51] | GA | GSHPs | SH | - | ✓ | - | Distributed control, multi-agent system | ✓ | - | - | Maximize the value of flexibility |
| [21] | MINLP | HP, flexible appliance loads, BESS, and solar PV system | SH | - | ✓ | - | Local control | Dynamic tariffs | ✓ | - | Minimize operation costs |
| [85] | LP | ASHP | SH and SC | - | ✓ | - | Local control | RTP | - | - | Minimize electricity costs |
| [22] | MILP | ASHP, electric water boiler, BESS, EV, and solar PV system | DHW and SH | - | ✓ | - | Local control | Fixed, exchange price, and TOU tariffs | ✓ | - | Maximize savings on electricity costs |
| [86] | LP, GA | GSHPs | SH and SC | - | ✓ | - | Centralized control by an load aggregator | Financial compensations | - | ✓ | Peak regulation and maximize the load aggregator revenue |

In the available literature, there are methods that also optimally coordinate the operation of devices across multiple buildings. Among these, the model described in [77] focuses on coordination in a community energy system (CES), involving ASHPs, EVs, energy storage systems for both homes and the community, flexible appliance loads, and solar PV systems, all of which are under TOU tariffs. The results demonstrate reductions in both electricity costs and carbon emissions. Similarly, [78] formulated a metaheuristic optimization model for centralized and decentralized coordination of ASHP and solar PV system operation across buildings. This model leverages DR with day-ahead and intraday electricity prices, showing that schedule-based optimization effectively minimizes electricity costs and surplus energy.

Further methods incorporate network objectives into optimization model formulation without the time-varying electricity prices of DR programs. The MILP model proposed in [80] coordinates multiple ASHPs in a residential district for minimizing peak power demand in local distribution networks. An LP model for controlling ASHPs is proposed in [46]. A central optimal power flow minimizes power losses while an individual controller performs peak shaving. The proposed convex optimization model proposed in [84] performs local and distributed coordination of GSHPs, showing a minimiza-tion of DR deviation and an increase in the GSHP COP. The optimization model proposed in [51] is based on a genetic algorithm (GA) and performs distributed control of multi-agent systems to regulate GSHPs. Another method [86] based on an LP model incorporated a financial compensation program. This method offered centralized control of GSHPs via a load aggregator that provided power peak regulation while maximizing the load aggregator revenue. Methods incorporating network objectives typically offer a coordination scheme for multiple devices, and the varying technical issues that arise can be addressed directly at the load nodes or with peak shaving, as a whole.

While optimization models based on scheduling offer control signals with time-series simulations, they usually lack real-time optimization. This characteristic is introduced by MPC methods, which deal with real-time control while retaining the formulation of an optimization model.

*C. Model Predictive Control*

The discussion on MPC for controlling HPs covers variants such as classic MPC, RMPC, SMPC, and AMPC, each adopting different approaches to the convexity and linearity of their optimization models. Table IV provides a comparative overview of the characteristics of these approaches for HP operation.

Most methods based on MPC aim to locally control HPs for customers enrolled in DR programs. The work in [87] proposed MPC based on a MILP model for controlling ASHPs for DHW heating and SH, along with a backup ERH. The pricing scheme is then established under dynamic TOU pricing from where the electricity generation is modeled. The results showed a reduction in electricity expenses for a customer. Another work [92] utilized MPC based on MIP modelling for controlling an ASHP for SH, minimizing electricity costs for energy billing through TOU tariffs. Also under TOU tariffs, the MPC in [99] controls an HP for SH, as well as a baseboard unit, EV, BESS, and solar PV system, showing the reduction in electricity costs with the proposed local controller. With an MPC method based on MILP, [90] controls a GSHP for DHW heating, SH, and SC, minimizing the total energy consumption and costs under RTP. The MPC in [93] controls a water-to-water HP for DHW heating and SH with an STC. The experimental approach showed a reduction in electricity costs for customers under RTP, including day-ahead prices and imbalance prices.

MPC algorithms have been proposed to optimize the operation of HPs and other devices for minimizing energy costs under different pricing schemes. The MPC algorithm in [91] controls an ASHP for SH and a GB to minimize heating costs under constant electricity prices, accounting for the cost of electricity and gas. The MPC in [101] controls a multi-source HP with an STC, minimizing electricity costs with consumption-based billing and local electricity rates, showing the results in a field test. Also with a field test approach, [3] proposed an MPC for operating an ASHP for SH with a backup ERH. The proposed approach minimized both energy consumption costs and peak demand costs since the energy billing accounts for constant electricity tariffs for consumption and peak power demand. The MPC in [37] operates an HP for DHW heating and SH, an EV, and a BESS, minimizing electricity costs and demand charges with DR utilizing a local energy market. The MPC in [97] minimizes operating costs and energy demand by controlling an ASHP for SH utilizing historical wholesale electricity prices.

Other methods formulate the control of HPs under DR programs with more robust MPC approaches to managing uncertainties using different algorithms. For instance, [57] and [95] proposed an SMPC algorithm controlling an ASHP for DHW heating and SH, minimizing electricity costs with day-ahead, intraday, and balancing markets. An SMPC developed in [23] controls an HP, a solar PV system, and an EV charger under time-varying electricity prices, minimizing electricity costs and reducing EV battery degradation costs. On the other hand, [58] developed an AMPC based on a quadratic programming (QP) model controlling an ASHP for DHW heating with an STC. The results showed a reduction in electricity costs with energy billing using RTP and TOU tariffs. The RMPC method proposed in [56] is based on a non-dominated sorting genetic algorithm (NSGA) which minimizes operating costs in a multi-objective approach. This model coordinates the operation of multiple GSHPs from buildings that are billed with TOU tariffs.

A range of models control HPs under DR programs also including solar PV generation. The MPC proposed in [88] controls an ASHP for DHW heating and SH and a solar PV system, maximizing self-consumption or minimizing electricity costs under constant electricity prices and time-varying electricity prices. The proposed MPC outperformed RBCs. Another MPC [94] controls an ASHP, a solar PV system, and a BESS under TOU tariffs, minimizing electricity costs. An MPC proposed in [100] defined the control of an HP, EWH, BESS, EV, and a solar PV system. The results show the minimization of electricity costs for RTP and demand limiting. Additionally, the paper adopted other objective functions for load shedding and shifting, and power tracking. The MPC from [55] operates an ASHP for DHW heating and SH with an STC, and a solar PV system. The model maximizes PV self-consumption and minimizes carbon emissions, electricity costs, and energy consumption with energy billing using electricity market hourly prices.

Some papers incorporate the coordination of HPs and solar PV systems with network objectives. The MPC based on the QP model designed in [89] controls an ASHP for DHW heating and SH with a backup ERH, minimizing electricity costs, environmental impact, and expansion of the generation capacity system, with TOU tariffs. The MPC method based on nonlinear programming (NLP) from [96] controls HPs for SH, solar PV systems, and EV chargers, minimizing electricity costs and EV battery degradation costs. This approach employs local MPC and centralized load flow analysis in an aggregator while utilizing TOU tariffs. Another work proposed in [98] controls the same devices also with local MPC and centralized load flow analysis in an aggregator, however, utilizing hourly spot electricity prices.

While MPC offers robust control, its dependence on accu-rate models highlights the need for more adaptable, model-free methods like RL. MPC presents a formulation based on constructing a model, which is widely adopted for controlling actual heating systems. However, the need for constructing an accurate model can make the approach vulnerable to sub-optimal solutions and limit its adaptability to changes in the system's dynamics. Therefore, model-free solutions introduce an alternative that does not depend on a model, aiming to interact with the environment and adapt to changes in the system's dynamics.

TABLE IV: Summary of MPC methods for HP operation.

| Ref. | Model | Devices | HP use | HP compressor speed Fixed | HP compressor speed Variable | Backup system | Energy flexibility focus DR | Energy flexibility focus PV | Energy flexibility focus Network | Objective |
|---|---|---|---|---|---|---|---|---|---|---|
| [87] | MPC, MILP | ASHP | DHW and SH | - | ✓ | ERH | Dynamic TOU pricing | - | - | Minimize electricity costs |
| [88] | MPC, convex QP | ASHP, TES, and solar PV panels | DHW and SH | - | ✓ | ERH | Constant electricity prices and time variable electricity prices | ✓ | - | Minimize electricity costs or maximize self-consumption |
| [89] | MPC, convex QP | ASHP and TES | DHW and SH | - | ✓ | ERH | TOU tariffs | ✓ | ✓ | Minimize electricity costs, environmental impact, and expansion of the generation capacity |
| [90] | MPC, MILP | GSHP | DHW, SH, and SC | ✓ | - | - | RTP | - | - | Minimize total energy consumption and cost |
| [91] | MPC, MILP | ASHP, gas boiler, and TES | SH | - | ✓ | GB | Constant electricity prices | - | - | Minimize heating costs (electricity and gas) |
| [92] | MPC, MIP | ASHP and TES | SH | - | ✓ | - | TOU tariffs | - | - | Minimize electricity costs |
| [93] | MPC | WSHP, TES, and STC | DHW and SH | - | ✓ | - | RTP, including day-ahead prices and imbalance prices | - | - | Minimize electricity costs |
| [94] | MPC | ASHP, solar PV panels, and BESS | SH | - | ✓ | ERH | Variable TOU tariffs | ✓ | - | Minimize electricity costs |
| [23] | SMPC, MINLP | HP, solar PV panels, and EV charger | SH | - | ✓ | - | Time-varying electricity price | ✓ | - | Minimize electricity costs and reduce EV battery degradation costs |
| [57] | SMPC | ASHP | DHW and SH | - | ✓ | - | Day-ahead, intraday, and balancing markets | - | - | Minimize electricity costs |
| [95] | SMPC | ASHP | DHW and SH | - | ✓ | - | Day-ahead, intraday, and balancing markets | - | - | Minimize electricity costs |
| [96] | MPC, NLP | HPs, solar PV panels, and EV chargers | SH | - | ✓ | - | TOU tariffs | ✓ | ✓ | MPC (Minimize electricity costs and EV battery degradation costs) |
| [97] | MPC, MILP | ASHP, and TES | SH | - | ✓ | - | Historical wholesale electricity prices | - | - | Minimize operating costs and energy demand |
| [98] | MPC, NLP | HPs, solar PV panels, and EV chargers | DHW and SH | - | ✓ | - | Hourly spot electricity prices | ✓ | ✓ | Minimize electricity costs and EV battery degradation costs |
| [58] | AMPC, QP | ASHP, STC, and TES | DHW | - | ✓ | - | RTP and TOU tariffs | - | - | Minimize electricity costs |
| [56] | RMPC, NSGA | GSHPs | SH | - | ✓ | - | TOU tariffs | - | - | Minimize operating costs |
| [99] | MPC | HPs, baseboard units, EVs, BESS, and solar PV panels | SH | - | ✓ | ERH | TOU tariffs | - | - | Minimize electricity costs |
| [100] | MPC, MILP | HP, EWH, BESS, EV, solar PV panels | DHW, SH, and SC | - | ✓ | - | RTP | ✓ | - | Minimize electricity costs for RTP and demand limiting. Other objective functions related to load shedding and shifting, and power tracking |
| [55] | MPC, MILP | ASHP, TES, STC, and solar PV panels | DHW and SH | - | ✓ | - | Electricity market hourly price | ✓ | - | Maximize PV self-consumption, minimize co2 emissions, minimize electricity costs, and minimize energy consumption |
| [101] | MPC, approximate MINLP | Multi-source HP, TES, and STC | DHW and SH | - | ✓ | - | Local electricity rates (field test) | - | - | Minimize electricity consumption |
| [3] | MPC, convex optimization model | ASHP | SH | - | ✓ | ERH | Constant electricity tariffs for consumption and peak power demand | - | - | Minimize peak power demand cost and electrical energy cost |
| [37] | MPC, MILP | HPs, TES, EVs, and BESS | DHW and SH | - | ✓ | - | Local energy market | - | - | Minimize electricity costs and demand charges |

*D. Reinforcement Learning-based Methods*

The discussion of the selected papers that developed RL-based algorithms for controlling HPs is supported by Table V, which provides a summary of these methods for HP operation. These methods include different RL variants and model-based MCTS, comparing them based on device characteristics and modeling, energy flexibility provision, and technical objectives.

Batch RL algorithms are employed to control HPs by training on predefined historical datasets. This enhances safety by avoiding the need for the agent to explore potentially unsafe actions during training, such as interacting with untested actions or state spaces. FQI [103] is used to control an HP and EWH to minimize electricity costs in dynamic pricing and day-ahead markets. In addition, [52] also applied FQI for controlling these devices under dynamic pricing, though it utilized simple on/off control instead of a variable-speed compressor, showing improved results compared to an RBC. A batch RL algorithm [104] for controlling an HP under TOU tariffs and day-ahead market conditions. The batch RL strategy [105] focused on managing an ASHP with local day-ahead electricity prices. Similarly, [10] employed FQI for controlling an HP with TOU tariffs. These methods demonstrated enhanced safety by using batch RL approaches while minimizing electricity costs across various time-varying electricity pricing schemes, such as TOU tariffs and day-ahead markets.

TABLE V: Summary of RL-based algorithms for HP operation.

| Ref. | Method [a] | Devices | HP use | HP compressor speed | | Backup system | Management approach | Energy flexibility focus | | | Objective |
|---|---|---|---|---|---|---|---|---|---|---|---|
| | | | | Fixed | Variable | | | DR | PV | Network | |
| [102] | RL | HP and solar PV system | DHW | ✓ | - | - | Local control | - | ✓ | - | Maximize self-consumption |
| [103] | Batch RL / FQI | HP and EWH | DHW and SH | - | ✓ | EWH | Local control | Dynamic pricing and day-ahead market | - | - | Minimize electricity costs |
| [104] | Batch RL | HP | SH | - | ✓ | - | Local control | TOU tariffs in day-ahead market | - | - | Minimize electricity costs |
| [105] | Batch RL | ASHP | SH | - | ✓ | - | Local control | Local day-ahead electricity prices | - | - | Minimize electricity costs |
| [52] | Batch RL / FQI | HP and EWH | DHW and SH | ✓ | - | EWH | Local control/DLC | Dynamic pricing | - | - | Minimize electricity costs |
| [106] | DRL with TL | HP and solar PV system | DHW | ✓ | - | - | Local control | TOU tariffs | ✓ | - | Minimize electricity costs and increase self-consumption |
| [44] | DRL / DQN | HP | DHW | ✓ | - | ERH | Local control | TOU tariffs | - | - | Minimize electricity costs |
| [107] | MCTS | HP | SH | - | ✓ | - | Local control | Spot prices | - | - | Minimize electricity costs |
| [24] | Safe RL based on PD-DDPG | HP, TES, mCHP, GB, and BESS | SH | - | ✓ | GB | Local control | Dynamic pricing | ✓ | - | Minimize electricity costs |
| [9] | MARL | HPs, EVs, flexible loads, and solar PV systems | SH | - | ✓ | - | Decentralized control | TOU tariffs | ✓ | ✓ | Minimize system costs |
| [108] | MPC-based RL | ASHP, TES, solar PV system, and BESS | SH | - | ✓ | - | Local control | Day-ahead spot market prices | ✓ | - | Minimize electricity costs |
| [109] | Recurrent SAC-DRL | HP, TES, and solar PV system | DHW and SH | - | ✓ | ERH | Local control | Dynamic spot electricity prices | ✓ | - | Minimize electricity costs |
| [110] | DRL | ASHP | SH | - | ✓ | - | Local control | Day-ahead spot electricity prices | - | - | Minimize electricity costs |
| [41] | Rainbow DRL | ASHP, DSHP, STC, and TES | Greenhouse heating | ✓ | - | - | Local control | Power-based demand charge and energy-based energy charge | - | - | Minimize electricity costs |
| [10] | FQI RL | HP | SH and SC | - | ✓ | - | Local control | TOU tariffs | - | - | Minimize electricity costs |
| [111] | MCTS | ASHP | SH | - | ✓ | - | Local control | Day-ahead price | - | - | Minimize electricity costs |
| [49] | SAC, TD3, PPO, and TRPO | ASHP, TES, and solar PV system | DHW and SH | - | ✓ | - | Local control | - | ✓ | - | Maximize self-consumption |
| [112] | Deep MARL | HPs, EVs with V2G, flexible loads, and solar PV systems | SH | - | ✓ | - | Centralized and decentralized cooperation | TOU tariffs | ✓ | ✓ | Minimize grid, voltage, and storage costs |

Other methods also focused on HP control under DR programs by utilizing DRL algorithms, which follow an online approach requiring the agent to continually interact with the environment during training. For instance, [44] proposed a DQN algorithm to control an HP for DHW heating alongside an electric-resistance backup system, both under a TOU tariff scheme. The study compared the performance of the RL-based approach with an MPC method, showing that the RL-based algorithm outperformed MPC in terms of electricity cost minimization. However, fine-tuning the MPC model remains critical to preserving its performance in these settings. Using a variation of a deep RL (DRL) algorithm, [110] utilized the PPO algorithm to control an ASHP for SH under day-ahead spot electricity prices. While these DRL methods have shown effectiveness in DR programs, they rely heavily on real-time interaction with the system.

In contrast, model-based RL offers an alternative by incorporating a predictive model of the environment, which reduces the need for extensive interaction with the actual system. Model-based methods, such as MCTS, have emerged as approaches for the energy management of HPs in SH, aiming to minimize electricity costs while learning the dynamics of the environment. An MCTS algorithm [107] controls an HP for SH under spot electricity prices. Similarly, [111] developed an MCTS method for controlling an ASHP, demonstrating the potential of this approach to minimize electricity costs in DR programs. However, a key disadvantage of MCTS compared to DRL and batch RL is that it can be computationally expensive and therefore less efficient for high-dimensional, continuous action spaces. These characteristics can present challenges when scaling the solution for the coordination of energy flexibility through HPs across multiple buildings in more complex real-world scenarios.

While most RL algorithms for DR programs focus on residential customers, some methods also address commercial or industrial customers, where power peak demand plays a significant role in energy billing. For instance, [41] applied a rainbow DRL for defining the control of ASHPs, DSHPs, a STC, for an end-user with greenhouse heating use. The algorithm minimized both power-based demand and energy-based charges by controlling these devices. The results showed that the algorithm was capable of minimizing the final electricity expenses while meeting the thermal demand. For demand charges, the algorithm must be able to hand the allocation of more than one compressor, for instance, making the overall peak demand lower for the energy billing period. Therefore, this DR program may also incorporate more than one compressor for HPs since more heating units may be under coordination.

The control of HPs is increasingly integrated into installations with DG systems like solar PV, where optimization of self-consumption and electricity cost reduction are key goals. For example, [24] developed a primal-dual deterministic policy gradient (PD-DDPG) algorithm to control a system comprising an HP, micro-combined heat and power (mCHP), GB, BESS, and solar PV under dynamic pricing schemes, indirectly managing self-consumption while minimizing electricity costs. Another approach [108] applied an MPC-based RL algorithm to control an ASHP, solar PV panels, and BESS with day-ahead spot market prices, combining both MPC and RL methods to further reduce electricity costs. A recurrent SAC-DRL algorithm [109] controls an HP for DHW heating and SH, and a solar PV system, showing the reduction in electricity expenses for dynamic spot electricity prices. Energy injection into the grid provides an alternative for selling excess generation during high production periods, while behind-the-meter (BTM) energy consumption can be used to supply heating demand in peak demand periods. However, limiting this injection was not considered, which might be a relevant factor in regions with high solar PV penetration.

RL can often demand significant computational time during the agent or multi-agent training process. An approach to mitigate this is transfer learning (TL), which improves the efficiency and speed of RL algorithms by leveraging knowledge from previously trained agents. For instance, [106] proposed an algorithm that uses TL to enhance the performance of a deep RL model for controlling an HP in DHW heating alongside a solar PV system. The method minimized electricity costs and increased self-consumption under TOU tariffs, demonstrating improved training efficiency. TL is particularly beneficial for RL algorithms applied to high-dimensional state and action spaces, which typically demand significant computational resources. For example, modeling the actions of a variable-speed compressor in an HP often involves continuous action spaces, which generally require more computational time for training agents compared to discrete action spaces, such as those representing on/off states.

While most of the selected methods provide control strategies for HPs that consider DR programs, a few focus on controlling HPs in conjunction with solar PV systems solely to maximize self-consumption, without integrating a DR program. For instance, [102] employed RL to operate an HP for DHW heating alongside a solar PV system, aiming to maximize self-consumption. In this approach, the HP utilizes on/off action states, and a backup controller is implemented to prevent violations of thermal comfort. Additionally, [49] evaluated other RL algorithms, including SAC, TD3, PPO, and TRPO, for controlling an ASHP used for DHW heating and SH in conjunction with a solar PV system, again focusing on maximizing self-consumption. This method outperformed an RBC used for comparison. Overall, these strategies emphasize local control with a customer-oriented objective of maximizing self-consumption, but they do not explicitly account for time-varying electricity prices.

Another set of papers has proposed algorithms that implement decentralized control of residential energy flexibility for a group of prosumers, utilizing multi-agent reinforcement learning (MARL) algorithms with network objectives. These studies have demonstrated the advantages of model-free approaches over centralized optimization models. For example, the MARL algorithm in [9] coordinates the operation of HPs for SH, EVs, flexible loads, and solar PV systems, minimizing system costs under TOU tariffs. This decentralized approach enhances scalability compared to centralized models. Additionally, another MARL algorithm [112] coordinates these devices under TOU tariffs but introduces network objectives to minimize network, voltage, and storage costs through both centralized and decentralized cooperation. These approaches are particularly beneficial for incorporating network objectives while also integrating DR programs, enabling end-users to reduce their electricity expenses.

While the selected methods primarily focus on local control strategies or the coordination of multiple devices within buildings, the effect of energy coordination on the distribution network is not always evaluated. A few studies, such as the RL-based algorithm in [112], do assess the impact on the grid. This network impact assessment remains an important objective, especially as some other papers examine the influence of HPs on the grid and the benefits of providing energy flexibility through the management of these thermostatically-controlled devices. While most RL approaches focus on optimizing individual device performance, understanding their broader impact on the grid is crucial for future applications. Therefore, the following section discusses the most relevant grid impact assessment studies that consider HPs, illustrating how the existing literature evaluates the benefits of energy flexibility offered by HPs in relation to network objectives.

## V. Grid Impact Assessment Studies

With the increasing integration of low-carbon technologies (LCTs), including HPs, their impact on existing distribution networks has become a prevalent field of study. Leveraging various modelling and simulation strategies, many emerging methods assess the installation of HPs on both existing and test networks, often combined with other LCTs like EVs, PV systems, and BESS. These studies, along with proposed control strategies, aim to show potential voltage and load violation mitigation, while identifying weaknesses in current distribution networks. In this context, Table VI provides a summary of the selected studies, emphasizing the devices considered, violation metrics, and HP flexibility strategies. This comparison specifically highlights the impact of HP adoption on distribution networks.

As shown in Table VI, devices are categorized into three types of HPs based on their heat source: ASHP, GSHP, and unspecified HPs, where the heat source is not specified and is considered a general heating system. Other categories include EV charging, solar PV generation, and other devices specifically considered in the references. On the violation metrics, these are related to line and transformer loading and voltage issues such as undervoltage, overvoltage, and voltage unbalance. Some studies further explore energy flexibility in HPs, extending beyond preheating, temperature setbacks, and equipment installation or upgrades. The discussion is organized around key themes, including probabilistic approaches, hosting capacity, and reinforcement requirements.

Some of the selected references offer a probabilistic approach for grid impact assessment. For instance, the impact of HPs on LV distribution networks is assessed in [113] using Monte Carlo simulations, assessing metrics such as voltage deviations, thermal loading, and transformer utilization with a specific emphasis on the influence of insulation, temperature, and auxiliary heating systems. The impact of HPs, solar PV systems, EVs, and micro CHP units on LV distribution networks is assessed in [114]. It quantifies voltage deviations, thermal loading, and transformer utilization across different penetration levels using a Monte Carlo-based probabilistic framework. The study in [115] evaluates how HPs affect distribution networks by employing a Monte Carlo method with Modelica-based simulations. It assesses voltage stability, thermal overloading of cables, and the influence of ASHP adoption on LV distribution feeders in rural and urban areas. Collectively, these studies aim to capture variability and uncertainty in grid impact assessments, particularly through Monte Carlo simulation frameworks.

Further studies also explore a probabilistic assessment of the impact of HPs in distribution networks. For instance, [116] evaluates how HPs affect a distribution network by employing Monte Carlo simulations and metamodeling to assess voltage stability, transformer loading, and the ability of LV distribution networks to accommodate HPs and PV systems across different penetration scenarios. Another work [121] assessed the impact of HPs on a distribution network by using a data-driven approach and Monte Carlo simulations to obtain metrics like load rate at distribution transformers, peak power fluctuations, and the likelihood of overloading in rural and suburban areas across different HP integration scenarios. The impact of HPs on a distribution network is evaluated in [120], which utilizes Monte Carlo simulations to assess voltage fluctuations and thermal limits. This approach determines the hosting capacity of LV distribution feeders and pinpoints potential congestion and voltage issues at various levels of HP penetration. These works offer different approaches for modelling heating load in combination with other residential devices.

Other studies focus on hosting capacity assessment methods for distribution networks. As an example, [123] evaluates the impact of HPs on LV distribution networks with a focus on voltage and thermal violations, occurrences of voltage drops, and network headroom. The study shows that the presence of HPs can cause voltage drops and thermal overloads. However, integrating optimized EV charging can help reduce these impacts and enhance the network's hosting capacity to accommodate more HPs without reduced reinforcement costs. The impact of HPs on a distribution network is analyzed in [118] through load flow simulations are utilized, assessing measures like voltage deviations, thermal utilization, and the necessity for grid reinforcement across different grid regions (urban, suburban, and rural) with varying levels of HP and EV penetration. Hosting capacity assessment offers distribution network planners information for defining the required expansion projects with the aim of meeting technical objectives.

The integration of HPs and other technologies poses challenges and opportunities for LV distribution networks, as demonstrated in recent studies. The impact of HPs, STC, TES systems, and solar PV systems on LV distribution networks is evaluated in [6], with a focus is on metrics such as voltage drops, which become more severe in winter but can be mitigated by the integration of STC and TES systems, thus aiding in maintaining voltages within acceptable limits. The work in [124] assesses the impact of HPs and solar PV systems on a distribution network by using integrated gas-electricity power flow analysis, evaluating key metrics such as substation utilization, voltage limit violations, and line congestion under various HP and rooftop solar PV adoption scenarios. The work in [125] assesses the impact of HPs alone on LV distribution networks through the use of geospatial data to model heat demand and assess its effects on network headroom, voltage unbalance, and thermal impacts. The impact of HPs and solar PV systems along EVs in Dutch LV distribution networks is evaluated in [5]. The analysis focuses on transformer overloading, line overloading, and voltage deviations. It is found that in winter, HPs lead to overloading and voltage deviations up to three times higher compared to other seasons. Finally, the methodology outlined in [126] evaluates the effects of HPs and EVs, on distribution network infrastructure, highlighting the significance of socio-economic factors in the modeling and decision-making processes for distribution network planning.

Some studies focus specifically on the impact of HPs on LV distribution networks, excluding the influence of other devices. For instance, [43] evaluates the impact of ASHPs on LV distribution networks, assessing metrics such as voltage and thermal violations, the percentage of customers experiencing voltage drops, and the length of network cables in thermal congestion, while analyzing the effectiveness of decentralized load-shifting strategies to manage these issues. The impact of HPs on a distribution network is evaluated in [117] by utilizing a predictive thermal relation model to synthesize HP electrical demand profiles based on temperature and time of day, assessing key metrics such as demand peaks, network headroom, and voltage stability under varying HP penetration scenarios. The proposed probabilistic model [119] evaluates how HPs affect a distribution network by simulating the electrical demand of HPs using real operational data and temperature relationships. It assesses key metrics such as demand peaks, network load, and the probability of surpassing thermal or voltage limits at various levels of HP penetration. The methodology described in [128] aims to forecast daily aggregated peak loads from HPs in the medium term in order to evaluate their effect on distribution networks' medium-voltage(MV)/LV transformers. These methods focus on the grid impact from HPs on distribution networks.

TABLE VI: Summary of grid impact assessment studies with HPs.

| Ref. | Devices | | | | | | Violation metrics | HP flexibility strategies |
|---|---|---|---|---|---|---|---|---|
| | ASHP | GSHP | Unspecified HP | EV | PV | other | | |
| [113] | ✓ | ✓ | - | - | - | - | Utilization index; number of customers with voltage problems; and thermal overloading | Favouring GSHP |
| [114] | ✓ | - | - | ✓ | ✓ | microCHP | Percentage of customers with voltage problems; loading level (max current/equipment capacity); and thermal overloading | - |
| [115] | ✓ | - | - | - | ✓ | - | Compared L-N min and max rms voltage of feeders to standard EN 50160; compared max rms current to manufacturer standards | - |
| [116] | ✓ | - | - | ✓ | - | - | Undervoltage (average feeder L-N rms voltage compared to limit); thermal overloading | - |
| [117] | ✓ | - | - | - | - | - | - | - |
| [118] | - | - | ✓ | ✓ | - | - | Coincidence Factor (simultaneity measure of peak demands for N customers); voltage deviations based on admissible range; overloading; thermal overloading | - |
| [119] | ✓ | - | - | - | - | - | - | - |
| [120] | - | - | ✓ | - | - | - | Average duration of voltage problems; average duration of congestion problems; overloading ("congestion problems/severity"); overvoltage ("voltage problems/severity"); undervoltage ("voltage problems/severity"); thermal overloading | - |
| [121] | - | ✓ | - | - | - | - | Overloading | - |
| [122] | ✓ | - | - | - | ✓ | - | Voltage deviation indicator; voltage unbalance indicator; feeder overloading indicator; transformer loading indicator | Feeder/equipment upgrades to accommodate HP integration |
| [123] | ✓ | ✓ | - | ✓ | ✓ | - | Undervoltage; overcurrent; feeder material restrictions; overloading; zonal power flow limit; thermal violations | HP headroom optimization; EV charging schedule that revolves around HP headroom |
| [43] | ✓ | - | - | - | - | - | Undervoltage; thermal violations | Preheating; Setback; Preheating + Setback |
| [124] | - | - | ✓ | - | ✓ | - | Overvoltage ("limit violations"); overloading | Installing various PV penetrations at various HP penetrations |
| [125] | - | - | ✓ | ✓ | - | - | Thermal overloading; overvoltage; undervoltage; voltage unbalance factor | - |
| [6] | - | - | ✓ | - | ✓ | BESS, TES systems | Overvoltage | Incorporating other tech like a BESS and TES systems to offset the HP load |
| [126] | - | - | ✓ | ✓ | - | - | Transformer overloading | - |
| [5] | ✓ | - | - | ✓ | ✓ | - | Line overloading; overvoltage; undervoltage | - |
| [127] | - | - | ✓ | ✓ | - | - | Transformer overloading; line overloading; undervoltage | - |
| [128] | - | - | ✓ | - | - | - | Transformer overloading | - |
| [129] | ✓ | - | - | - | ✓ | - | Voltage deviation; voltage unbalance; feeder overloading; Transformer Overloading | - |

More recent published references focused on incorporating utility-based reinforcement of distribution networks into the grid impact assessment studies. Specifically, in [122], the impact of incorporating HPs and solar PV systems into LV grids is evaluated with a focus on measures such as voltage deviations, voltage unbalance, overloading of feeders and transformers, and the costs associated with reinforcement. The work in [127] assesses the impact of EV charging and HPs on LV distribution networks, identifying potential constraints, and providing a framework for grid reinforcement to accommodate future energy demands. In [129], the impact of integrating residential HPs and solar PV systems into LV distribution networks is analyzed from a techno-economic perspective. The focus is on the expenses associated with grid reinforcements and the advantages of enhanced insulation in mitigating these costs. These studies focusing on reinforcement criteria offer information for grid expansion studies, including potential required investment costs, considering grids with increasing adoption of HPs.

## VI. CONCLUSIONS

This paper presents a comprehensive review of control strategies for HPs aimed at energy management, organized into broader categories based on their energy flexibility strategies. Most of the reviewed methods concentrated on achieving energy flexibility by controlling devices within DR programs, primarily aiming to minimize electricity costs under time-varying pricing schemes. TOU tariffs, dynamic pricing, and day-ahead market prices were widely applied in residential settings. However, the integration of HP control to address additional considerations, such as demand charges, was predominantly explored in commercial applications.

Methods incorporating solar PV generation often focus on maximizing energy self-consumption, either explicitly by including it in the objective or reward function, or indirectly through the minimization of electricity costs via heating demand control. Systems that included setups with BESS further optimized electricity costs for heating. However, most studies did not account for battery degradation, which can significantly affect the overall cost savings.

While energy flexibility strategies with a focus on DR programs and support for the integration of solar PV generation generally focus on the local control of individual devices, methods targeting network objectives predominantly employed coordinated management approaches, involving devices across multiple households. These methods addressed objectives like minimizing peak power demand, reducing power losses, mitigating voltage issues, or enhancing grid hosting capacity. By explicitly modeling utilities' technical goals, these methods offer an alternative to solely relying on DR programs, which are heavily influenced by variations in end-users' energy consumption behavior.

To highlight the need for energy flexibility strategies through HP operation for energy management purposes, we provided a review of distribution grid impact studies for HPs. While many studies focus on assessing devices and violation metrics, as well as the benefits of incorporating HP load control into grid-level metrics, there remains a significant gap in exploring advanced control approaches. The absence of methods such as MPC and RL within these studies limits the potential for achieving energy flexibility.

Some of these studies further assess the benefits of flexibility strategies including HP load control into the distribution grid level metrics. However, the application of more advanced control methods, such as MPC and RL algorithms, was not explored in these studies, despite their potential to enhance energy flexibility and benefit distribution networks.

Building on the identified research gaps, we propose the following directions to enhance HP control strategies for energy management.

## A. Recommendations for Future Research

The use of backup heating systems such as GB or ERH, demonstrates the need for testing scenarios that evaluate device control strategies in extreme cold-weather regions. In such regions, HPs often require backup heaters to maintain thermal comfort during winter. These scenarios are crucial for assessing the effectiveness of energy flexibility strategies, ensuring validation not only in terms of achieving electricity cost minimization but also in maintaining indoor ambient and hot water temperatures within acceptable limits.

From the perspective of commercial and industrial customers with HPs, control strategies for demand charge schemes are particularly relevant. Future research could expand on this by considering demand charge schemes incorporating customer preferences such as EV charging schedules. However, there are limitations to incorporating such preferences, particularly the influence of non-controllable loads, which could affect peak demand and overall electricity costs.

Advancements in artificial intelligence, particularly in TL, have the potential to reduce the training time of RL agents. For example, TL was utilized to accelerate training by leveraging previously trained agents to control an HP for DHW heating in conjunction with a solar PV system. Future research could extend this approach to control HPs with both SH and DHW heating functions, incorporating considerations for backup heating systems. The training process for RL becomes more resource-intensive in MDPs with high-dimensional state and action spaces, leading to significant computational demands. For instance, modeling the actions of a variable-speed compressor in an HP often involves continuous action spaces, which generally require more computational time for training agents compared to discrete action spaces, like those representing on/off states.

Future research could explore coordinating the operation of HPs for both SH and DHW heating, along with other residential devices, to address technical objectives at the primary distribution network level. This study could also investigate other DR programs, including dynamic pricing, with aggregators facilitating the coordination of groups of residential loads.

Among the reviewed grid impact assessment studies of HPs, some also evaluated the benefits of energy flexibility strategies, including HP load control, on distribution grid-level metrics. While these studies primarily focused on RBCs designed for coordination purposes, they did not address more advanced methods, such as real-time optimization using MPC or RL-based algorithms, in their grid impact analyses. Incorporating these advanced control methods could serve to identify potential limitations, such as the impact of varying user participation levels and the cost-benefit trade-offs of implementing these methods.

While the authors provide recommendations for future research, the summary tables highlight additional characteristics of the selected methods. These include both approaches to HP control for congestion management and grid impact assessment studies of HPs.

ACKNOWLEDGMENT

This work was supported by Mitacs through the Mitacs Accelerate program.

ABBREVIATIONS

| | |
|---|---|
| ASHP | Air-source heat pumps |
| BESS | Battery energy storage systems |
| CHP | Combined heat and power |
| COP | Coefficient of performance |
| CPP | Critical peak pricing |
| DDPG | Deep deterministic policy gradient |
| DERs | Distributed energy resources |
| DHW | Domestic hot water |
| DLC | Direct load control |
| DQL | Double Q-learning |
| DQN | Deep Q-network |
| DR | Demand response |
| DRL | Deep reinforcement learning |
| DSHP | Dual-source heat pumps |
| DSM | Demand-side management |
| DSO | Distribution system operator |
| EMS | Energy management systems |
| EV | Electric vehicles |
| EWH | Electric water heating |

| FQI | Fitted Q-iteration |
| --- | --- |
| GSHP | Ground-source heat pumps |
| HEMS | Home energy management systems |
| HP | Heat pumps |
| HVAC | Heating, ventilation, and air conditioning |
| LCT | Low-carbon technologies |
| LV | Low-voltage |
| MARL | Multi-agent reinforcement learning |
| MCTS | Monte Carlo tree search |
| MDP | Markov decision process |
| MILP | Mixed-integer linear programming |
| MINLP | Mixed-integer nonlinear programming |
| MPC | Model predictive control |
| MV | Medium-voltage |
| PD-DDPG | Primal-dual deterministic policy gradient |
| PPO | Proximal policy optimization |
| PV | Photovoltaic |
| QP | Quadratic programming |
| RBC | Rule-based controllers |
| RL | Reinforcement learning |
| RMPC | Robust model predictive control |
| RTP | Real-time pricing |
| SAC | Soft actor-critic |
| SC | Space cooling |
| SH | Space heating |
| SMPC | Sthocastic model predictive control |
| STC | Solar thermal collector |
| TCL | Thermostatically-controlled loads |
| TD3 | Twin delayed deep deterministic policy gradient |
| TES | Thermal energy storage |
| TL | Transfer learning |
| TOU | Time-of-use |
| TRPO | Trust region policy optimization |